\documentclass[aps,prb,showpacs,twocolumn,superscriptaddress,floatfix]{revtex4}
\usepackage{graphicx,color,multirow,textcomp}
\begin{document}
\pdfoutput=1
\title{Mesoscopic persistent currents in a strong magnetic field}
\author{Eran Ginossar}
\affiliation{Department of Physics, Yale University, New Haven, Connecticut 06520, USA}
\author{Leonid I.\ Glazman}
\affiliation{Department of Physics, Yale University, New Haven, Connecticut 06520, USA}
\affiliation{Department of Applied Physics, Yale University, New Haven, Connecticut 06520, USA}
\author{Teemu Ojanen}
\affiliation{Dahlem Center for Complex Quantum Systems and Fachbereich Physik, Freie Universit\"at Berlin, 14195 Berlin, Germany}
\author{Felix von Oppen}
\affiliation{Dahlem Center for Complex Quantum Systems and Fachbereich Physik, Freie Universit\"at Berlin, 14195 Berlin, Germany}
\author{William E.\ Shanks}
\affiliation{Department of Physics, Yale University, New Haven, Connecticut 06520, USA}
\author{Ania C.\ Bleszynski-Jayich}
\affiliation{Department of Physics, Yale University, New Haven, Connecticut 06520, USA}
\author{J.\ G.\ E.\ Harris}
\affiliation{Department of Physics, Yale University, New Haven, Connecticut 06520, USA}
\affiliation{Department of Applied Physics, Yale University, New Haven, Connecticut 06520, USA}

\date{\today}
\begin{abstract}
Recent precision measurements of mesoscopic persistent currents in normal-metal rings rely on the interaction between the magnetic moment generated by the current and a large applied magnetic field. Motivated by this technique, we extend the theory of mesoscopic persistent currents to include the effect of the finite thickness of the ring and the resulting penetration of the large magnetic field. We discuss both the sample-specific typical current and the ensemble-averaged current which is dominated by the effects of electron-electron interactions. We find that the magnetic field strongly suppresses the interaction-induced persistent current and so provides direct access to the independent-electron contribution. Moreover, the technique allows for measurements of the entire distribution function of the persistent current. We also discuss the consequences of the Zeeman splitting and spin-orbit scattering, and include a detailed and quantitative comparison of our theoretical results to experimental data.
\end{abstract}
\maketitle

\section{Introduction}

Persistent currents in normal-metal rings threaded by an Aharonov-Bohm flux constitute a paradigm of quantum-coherence effects in the thermodynamic properties of mesoscopic systems. While the history of persistent currents dates back to the early days of quantum mechanics \cite{Hund38} and of superconductivity,\cite{superconductivity} they were studied intensively starting with the seminal paper by Buttiker, Imry, and Landauer.\cite{Buttiker83}

Most experiments to date detected persistent currents using SQUIDs (superconducting quantum interference device) as magnetometers. A different technique was recently developed by Bleszynski-Jayich {\em et al.}\cite{Harris09} which is much more sensitive and allows for precision measurements with lower back action and over a wider range of magnetic fields. The high-precision cantilever torque magnetometer relies on the interaction of the magnetic moment associated with the persistent current and a large applied magnetic field. This interaction shifts the resonance frequency of a microcantilever on which the rings are located. Measurements of the frequency shift allow one to extract the persistent current quantitatively.

This paper extends the existing theory of mesoscopic persistent currents to include a large applied magnetic field. We focus on metallic samples with diffusive electron dynamics for which the applied magnetic fields are {\it non}-quantizing. Our results hold for both normal-metal rings as well as rings made of nominally superconducting materials (provided that the magnetic field significantly exceeds the superconducting critical field $H_{c2}$) and include the effects of spin, namely Zeeman splitting and spin-orbit scattering.

Within an independent-electron model, the flux-periodic persistent current is strongly sample specific with both magnitude and sign depending on the details of the disorder configuration and the geometry of the ring. As a result, its ensemble average $\langle I\rangle$ is, even in the canonical ensemble,\cite{Altshuler91,Schmid91,Oppen91} small compared to its second moment $\langle I^2\rangle$ \cite{Cheung89,Riedel93} so that the latter describes the typical magnitude of the persistent current of an individual ring. The typical persistent current is $\phi_0$-periodic and, in a diffusive metallic ring, has an amplitude of the order of $e/\tau_D$, where $\tau_D$ denotes the diffusion time of an electron around the ring.

The disorder-averaged persistent current is dominated by the contribution of electron-electron interactions \cite{Ambegaokar90} and is $\phi_0/2$-periodic. In a normal-metal ring, it is of the order of $\lambda(e/\tau_D)$, where $\lambda$ is an effective electron-electron coupling constant. While $\lambda$ is of order unity in lowest order perturbation theory, higher-order contributions are expected to reduce its magnitude significantly. \cite{Altshuler81,Eckern91} In rings made of superconducting materials, a related mechanism leads to a current due to superconducting fluctuations above the critical temperature $T_c$.\cite{Eckern90, Oppen92}

These theoretical expectations have been tested in several experiments, including metallic, \cite{Levy90, Chandrasekhar91, Jariwala01, Moler09, Harris09} semiconducting,\cite{Mailly93} as well as superconducting\cite{Moler07} rings. While results of early experiments with metallic rings were in apparent strong disagreement with theoretical predictions, a more recent SQUID-based experiment \cite{Moler09} yielded data reasonably close to theory. Finally, the measurement of the typical persistent current reported in Ref.\ \onlinecite{Harris09} agrees, without any adjustable parameters and over a wide range of experimental variables, with the predictions of the model of noninteracting diffusive electrons described here. We also note in passing that there is a closely related set of works, both experimental and theoretical, which explores the magnetic response of singly-connected mesoscopic systems, see e.g., Refs.\ \onlinecite{Levy93, Oppen94, Ullmo95, Ullmo98}.

Persistent currents have also motivated a multitude of further theoretical considerations. Among other results, it was suggested that the persistent current is highly sensitive to a variety of subtle effects, including the coupling of the ring to its electromagnetic environment \cite{Yudson93, Aronov93} as well as magnetic impurities within the ring.\cite{Bary08, Schwiete09} This indicates that accurate measurements and understanding of persistent currents in various settings would address a number of interesting questions in many-body condensed matter physics.

This paper is organized as follows. In Sec.\ \ref{periodicity}, we discuss the flux dependence of the persistent current. Sec.\ \ref{independent} discusses the effects of the strong magnetic field on the persistent current within the independent-electron model, including the effects of the Zeeman energy and spin-orbit scattering. Sec.\ \ref{interaction} focuses on the interaction contribution to the persistent current. Sec.\ \ref{datacomparison} contains a detailed comparison between our theoretical results and the experimental data of Ref.\ \onlinecite{Harris09}. We conclude in Sec.\ \ref{conclusions}.

\section{Flux periodicity of the persistent current}
\label{periodicity}

Conventionally, persistent currents are discussed in the limit of a pure Aharonov-Bohm flux threading the ring. In this case, gauge invariance implies flux periodicity,\cite{Byers}
\begin{equation}
  I(\phi)=I(\phi+\phi_0),
\end{equation}
where the period is given by the flux quantum $\phi_0=h/e$, and time-reversal invariance gives the relation
\begin{equation}
  I(\phi)=-I(-\phi).
\end{equation}
As a result, the persistent current vanishes at integer and half-integer multiples of the flux quantum and can be expressed as a Fourier series $I(\phi)=\sum_{p=1}^\infty I_p \sin(2\pi p \phi/\phi_0)$.

It is instructive to deduce the consequences of this Fourier decomposition for the current-current correlation function
\begin{equation}
  C(\phi,\phi')=\langle I(\phi)I(\phi')\rangle.
\end{equation}
Here, $\langle\ldots\rangle$ denotes a disorder average. We anticipate that the Fourier components $I_p$ are mutually uncorrelated, i.e., $\langle I_p I_{p'}\rangle = \langle I_p^2\rangle \delta_{pp'}$. Then, we obtain
\begin{widetext}
\begin{eqnarray}
   C(\phi,\phi') &=& \sum_{p=1}^\infty \langle I_p^2\rangle \sin(2\pi p \phi/\phi_0) \sin(2\pi p \phi'/\phi_0)
      = \sum_{p=1}^\infty \frac{\langle I_p^2\rangle}{2} \left\{ \cos[2\pi p (\phi-\phi')/\phi_0] - \cos[2\pi p
      (\phi+\phi')/\phi_0]\right\}.
\end{eqnarray}
\end{widetext}
Within the diagrammatic approach to diffusive electronic systems, the two terms depending on $(\phi-\phi')$ and $(\phi+\phi')$ have immediate interpretations as the diffuson and cooperon contributions, respectively.\cite{Riedel93} Both contributions are of the same magnitude but depend differently on the magnetic flux.

In the presence of an additional large magnetic field $B$ penetrating the metal ring, one expects that the cooperon contribution is strongly suppressed. This leads to a change in the flux dependence of $C(\phi,\phi^\prime)$ which can also be obtained directly from symmetry considerations. While gauge invariance and hence the flux periodicity persist, the additional magnetic field changes the time-reversal relation into $I(B,\phi)=-I(-B,-\phi)$. As a result, the current is no longer odd in the Aharonov-Bohm flux $\phi$ alone, and the Fourier series takes the more general form
(at fixed $B$)
\begin{equation}
  I(\phi) = \sum_{p=1}^\infty \{I^{(+)}_p \cos(2\pi p\phi/\phi_0)+I^{(-)}_p \sin(2\pi p\phi/\phi_0)\}.
  \label{FourierDecomposition}
\end{equation}
If we again anticipate that the Fourier components are mutually uncorrelated,
\begin{eqnarray}
  \langle I^{(\pm)}_p I^{(\pm)}_{p'}\rangle &=& \langle [I^{(\pm)}_p]^2\rangle \delta_{pp'}
  \label{Correlators1}\\
  \langle I^{(+)}_pI^{(-)}_{p'}\rangle &=& 0
  \label{Correlators2}
\end{eqnarray}
and that, moreover, $\langle [I^{(+)}_p]^2\rangle = \langle [I^{(-)}_p]^2\rangle $, we find
\begin{widetext}
\begin{eqnarray}
   C(\phi,\phi')
   &=&
    \sum_{p=1}^\infty \langle [I^{(+)}_p]^2\rangle [ \sin(2\pi p \phi/\phi_0) \sin(2\pi p \phi'/\phi_0) + \cos(2\pi p \phi/\phi_0) \cos(2\pi p \phi'/\phi_0) ]
  \nonumber  \\ & =&
   \sum_{p=1}^\infty \langle [I^{(+)}_p]^2\rangle \cos[2\pi p (\phi-\phi')/\phi_0] .
   \label{HarmonicContent}
\end{eqnarray}
\end{widetext}
In agreement with expectations, our analysis implies that in the presence of a large magnetic field $B$, the current-current correlation function has the flux dependence of a diffuson contribution. Note that the magnitude of the persistent current, $\langle I^2(\phi) \rangle$, becomes independent of flux.  As a special case, this also implies that the persistent current can be nonzero at zero flux.

It is interesting to note that in the presence of a large magnetic field, the flux dependence of the persistent current can also be written as
\begin{equation}
 I(\phi) = \sum_{p=1}^\infty I_p \cos(2\pi p \phi/\phi_0 - \alpha).
 \label{RandomPhase}
\end{equation}
Comparing with Eq.\ (\ref{FourierDecomposition}) yields the identities $I_p^{(+)} = I_p \cos\alpha$ and $I_p^{(-)} = I_p \sin\alpha$. Then, we automatically reproduce Eqs.\ (\ref{Correlators1}) and (\ref{Correlators2}) by assuming that the phase offset $\alpha$ has a uniform distribution over the disorder ensemble. This also yields the relation $\langle I_p^2\rangle = 2\langle [I^{(\pm)}_p]^2\rangle $.

In the next section, we verify these flux dependencies explicitly within the model of diffusive non-interacting electrons.

\section{Independent-electron contribution}
\label{independent}

\subsection{Current-current correlation function}

\begin{figure}[t]
                \includegraphics[width=0.8\columnwidth]{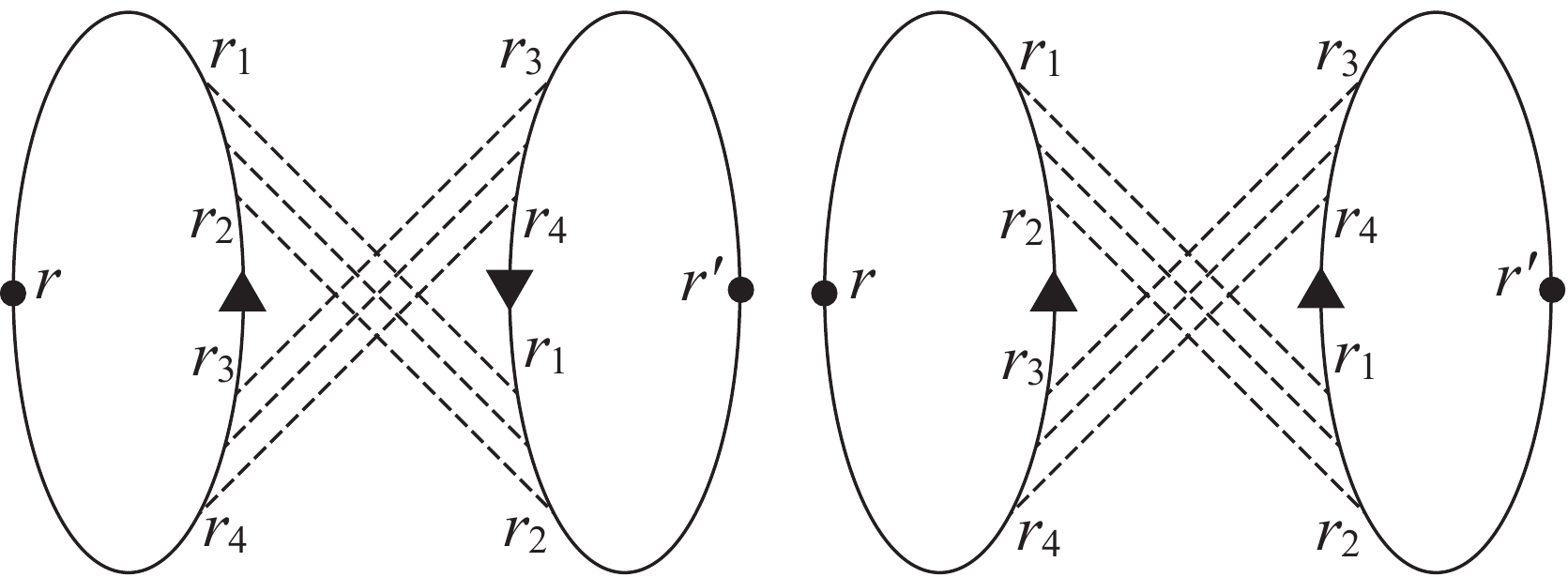}
\caption{(a) Diffuson and (b) cooperon diagrams for the autocorrelation function of the density of states. Full lines represent electronic Green functions, dashed lines correspond to disorder scattering.
\label{diagrams}}
\end{figure}

The persistent current is obtained as the flux-derivative of the thermodynamic potential
\begin{equation}
  I=-\frac{\partial\Omega}{\partial \phi}.
\end{equation}
For non-interacting electrons, the (grand-canonical) thermodynamic potential $\Omega$ can be expressed as
\begin{equation}
  \Omega(\mu,{\bf B}) = -T \int dE\, \nu(E,{\bf B}) \ln[1+e^{-\beta(E-\mu)}]
\end{equation}
in terms of the density of states $\nu(E,{\bf B})$. Here, $\beta=1/T$ denotes the inverse temperature. (We use units $k_B=1$ and $\hbar=1$.) Here, the magnetic field ${\bf B}$ includes both the Aharonov-Bohm flux $\phi$ threading the ring and the magnetic field penetrating the ring. For definiteness, we will from now on decompose the full magnetic field into a pure Aharonov-Bohm contribution and an in-plane field $B_\parallel$ penetrating the ring. Accordingly, we will drop the vector nature of ${\bf B}$ in the following although it should be kept in mind that in principle, the persistent current is not an isotropic function of magnetic field.

The thermodynamic potential $\Omega(\mu,B)$ at finite temperature can be related to its zero-temperature limit
\begin{equation}
 \Omega_0(\mu,B) = \int_{-\infty}^\mu dE (E-\mu) \nu(E,B)
\end{equation}
as
\begin{equation}
  \Omega(\mu,B) = \int_{-\infty}^\infty dE \left(-\frac{\partial f_\mu(E)}{\partial E}\right) \Omega_0(E,B),
\end{equation}
in terms of the Fermi-Dirac distribution $f_\mu(E)$. Thus, the current-current correlation function $C_I(B,B')=\langle I(B)I(B')\rangle$ takes the form
\begin{widetext}
\begin{eqnarray}
  C_I(B,B') &=& \int dE\, dE' \left(-\frac{\partial f_\mu(E)}{\partial E}\right) \left(-\frac{\partial f_\mu(E')}{\partial E'}\right) \frac{\partial^2}{\partial \phi\partial \phi'} \,\langle \Omega_0(E,B) \Omega_0(E',B') \rangle \nonumber\\
   &=&  \int d\epsilon \,\partial_\epsilon^2 \left(\frac{\epsilon}{1-\exp(-\beta\epsilon)}\right) \frac{\partial^2}{\partial \phi\partial \phi'} \,\langle\Omega_0(E,B)\Omega_0(E',B')\rangle \nonumber\\
   &=&  \int d\epsilon \,\partial_\epsilon^2 \left(\frac{\epsilon}{1-\exp(-\beta\epsilon)}\right)\,C_I^{(0)} (E,B;E',B')
   \label{FiniteTemperature}
\end{eqnarray}
Here, we used in the second identity that the correlator depends only on the energy difference $\epsilon=E-E'$ so that we can perform the integral over the sum $\sigma=E+E'$. Thus, we are led to consider the zero-temperature autocorrelation function $C_I^{(0)} (E,B;E',B') = \langle I(E,B) I(E',B')\rangle$ of currents at different chemical potentials $E$ and $E'$ as well as fields $B$ and $B'$.

Within a model of non-interacting, diffusive electrons, the calculation of
\begin{equation}
C_I^{(0)} (E,B;E',B') = \int_{-\infty}^E dE_1 \int_{-\infty}^{E'} dE_2 (E_1-E)(E_2-E')\frac{\partial^2}{\partial \phi\partial \phi'} \langle \nu(E_1,B)\nu(E_2,B')\rangle
\end{equation}
starts from the familiar diagrams in Fig.\ \ref{diagrams} for the disorder-averaged autocorrelation function of the density of states.\cite{Shklovskii} Note that {\it both} the diffuson and the cooperon diagram contribute to the persistent current. The diffuson diagram depends on the difference ${\bf A}_-={\bf A}-{\bf A}'$ of the magnetic vector potentials, the cooperon diagram on the sum ${\bf A}_+={\bf A}+{\bf A}'$. Performing the integrations over the fast Green-function arguments in the diagrams of Fig.\ \ref{diagrams}, one arrives at the expression
\begin{eqnarray}
 C_I^{(0)}(E,B;E',B') = \frac{1}{2\pi^2}\frac{\partial^2}{\partial\phi\partial \phi'} {\rm Re} \sum_{\pm}\int_0^\infty
  d\sigma  \int_{-2\sigma}^{2\sigma} d\epsilon \left[\sigma^2 - \frac{\epsilon^2}{4} \right]
  {\rm Tr}\left(\frac{1}{-D[\nabla-{ie}{\bf A}_\pm]^2 + i(\epsilon+E-E')}\right)^2.
\end{eqnarray}
Rewriting the square of the diffusion pole as a derivative with respect to $\epsilon$ and integrating by parts yields
\begin{eqnarray}
 C_I^{(0)}(E,B;E',B') = -\frac{1}{4\pi^2}\frac{\partial^2}{\partial\phi\partial \phi'}  \sum_{\pm}\int_0^\infty
  d\sigma {\rm Im} \int_{-2\sigma}^{2\sigma} d\epsilon\, \epsilon
  {\rm Tr}\left(\frac{1}{-D[\nabla-{ie}{\bf A}_\pm]^2 + i(\epsilon+E-E')}\right).
\label{CurrentThroughDifusion}
\end{eqnarray}
\end{widetext}
Here, $D$ denotes the diffusion constant and we limit attention to spinless systems. (Effects of spin will be discussed separately in Sec.\ \ref{spin}.)

In Eq.\ (\ref{CurrentThroughDifusion}), the trace is over a space of wavefunctions $\psi$ satisfying the condition
\begin{equation}
 \left. {\bf \hat n}\cdot[\nabla-{ie}{\bf A}_\pm]\psi\right|_\Sigma =0.
 \label{BoundaryCondition}
\end{equation}
at the surface $\Sigma$ of the metallic ring. (${\bf \hat n}$ denotes denotes a unit vector normal to the surface.) In general, this boundary condition makes the evaluation of Eq.\ (\ref{CurrentThroughDifusion}) a tedious problem.

To simplify this problem, we use a model in which the in-plane magnetic field is taken to be of constant magnitude and to point along the azimuthal direction around the ring. While this toroidal-field model is clearly different from experimental realizations, we expect that it gives a qualitatively and, for certain quantities, even quantitatively correct account of the consequences of a large magnetic field penetrating the ring. Specifically, we expect that the predictions for the correlation field $B_c$ are parametrically correct while the numerical prefactor would reflect the particular field configuration. At the same time, predictions for the typical current amplitude will be quantitatively correct because the large in-plane field drops out of the final expressions.

Some considerations for more general field configurations are collected in an Appendix.

\subsection{Toroidal magnetic field}
\label{ToroidalMagneticField}

The simplification of the toroidal-field model derives from the fact that in this case, the eigenvalue problem
\begin{equation}
-D[\nabla-{ie}{\bf A}_\pm]^2 \psi = {\cal E}\psi
\label{diffuson}
\end{equation}
together with the boundary condition in Eq.\ (\ref{BoundaryCondition}) can be solved by separation of variables. Let us consider a ring defined as a cylinder of length $L$ (along the $z$-direction) and radius $R$ (in the $x-y$-plane) with periodic boundary conditions in the $z$-direction. The total vector potential ${\bf A}$ is a sum of the Aharonov-Bohm contribution ${\bf A}_\perp= (\phi/L) {\bf \hat z}$ describing the flux threading the ring and the vector potential ${\bf A}_\parallel = (B_\parallel/2) {\bf \hat z}\times{\bf r}$ of the in-plane magnetic field penetrating the ring. Then, the eigenvalue problem in Eq.\ (\ref{diffuson}) separates with $\psi(x,y,z) = \chi(x,y)\exp(ikz)$ where
\begin{equation}
 {\cal E} = E_c (n-\varphi_\pm)^2 + \epsilon_\perp
\end{equation}
with $n=0,\pm 1,\pm2 \ldots$ and
\begin{equation}
-D [ (\partial_x -\frac{ieB}{2}y)^2 + (\partial_y +\frac{ieB}{2}x)^2 ] \chi = \epsilon_\perp \chi.
\label{DiffusonEquation}
\end{equation}
Here, we defined the Thouless energy
\begin{equation}
  E_c= \frac{4\pi^2  D}{L^2}
\end{equation}
and the dimensionless flux variable $\varphi_\pm = \phi_\pm/\phi_0$. Note that in order not to introduce unnecessary numerical prefactors into equations, this definition of the Thouless energy differs by a factor of four from the definitions employed in Refs.\ \onlinecite{Harris09} and \onlinecite{Riedel93}.

Inserting these eigenvalues into Eq.\ (\ref{CurrentThroughDifusion}), we find
\begin{widetext}
\begin{eqnarray}
 C_I^{(0)}(E,B;E',B') = -\frac{1}{4\pi^2}\frac{\partial^2}{\partial\phi\partial \phi'}  \sum_{\pm}\sum_{\epsilon_\perp}\sum_n \int_0^\infty
  d\sigma {\rm Im} \int_{-2\sigma}^{2\sigma} d\epsilon\, \epsilon
  \frac{1}{E_c (n-\varphi_\pm)^2 + \epsilon_\perp + i(\epsilon+E-E')}.
\end{eqnarray}
Performing the sum over $n$ by Poisson summation and measuring all energy variables in units of the Thouless energy, one obtains
\begin{eqnarray}
 C_I^{(0)}(E,B;E',B') = -\frac{E_c^2}{2\pi}\frac{\partial^2}{\partial\phi\partial \phi'}  \sum_{\pm}\sum_{\epsilon_\perp}\sum_{p=1}^\infty \cos(2\pi p\varphi_\pm)\int_0^\infty
  d\sigma {\rm Im} \int_{-2\sigma}^{2\sigma} d\epsilon\, \epsilon
 \frac{\exp(-2\pi p \sqrt{\epsilon^\pm_\perp + i(\epsilon+E-E')})}{\sqrt{\epsilon^\pm_\perp + i(\epsilon+E-E')}}.
\end{eqnarray}
The integrals over $\epsilon$ and $\sigma$ can be readily done to yield
\begin{eqnarray}
 C_I^{(0)}(E,B;E',B') &=& -\frac{8E_c^2}{\pi}\frac{\partial^2}{\partial\phi\partial \phi'}  \sum_{\pm}\sum_{\epsilon_\perp}\sum_{p=1}^\infty \cos(2\pi p\varphi_\pm) F_p(z_\pm)
\label{ZeroTemperature}
\end{eqnarray}
where $z_\pm=[\epsilon^\pm_\perp + i(E-E')]/E_c$ and where we defined the function
\begin{equation}
F_p(z) = {\rm Re} \left[ \left( \frac{3}{(2\pi p)^5} + \frac{3\sqrt{z}}{(2\pi p)^4} + \frac{z}{(2\pi p)^3} \right) e^{-2\pi p\sqrt{z}}\right]
\label{functioFp}
\end{equation}
\end{widetext}
We note that this result is valid for spinless fermions. Effects of spin will be discussed below in Sec.\ \ref{spin}.

It is interesting to compare the result in Eq.\ (\ref{ZeroTemperature}) with the corresponding correlation function for the conductance fluctuations of a metallic ring.\cite{Aronov87,Lee87} Indeed, the flux-sensitive contributions to the correlation function of the conductance at different magnetic fields differ from our result for the persistent current (apart from an overall prefactor) only by the preexponential factor in the function $F_p(z)$.

In the absence of the in-plane magnetic field, we need to retain only the lowest transverse eigenvalue $\epsilon_\perp^\pm =0$ to exponential accuracy in $2L/R$. Then, we find
\begin{equation}
\langle I(\phi)I(\phi')\rangle = \frac{6E_c^2}{\pi^4 \phi_0^2}\sum_{p=1}^\infty \frac{1}{p^3}\sin(2\pi p \varphi)\sin(2\pi p \varphi')
\label{IInoSpin}
\end{equation}
for the current-current correlation, which reproduces the result obtained in Ref.\ \onlinecite{Riedel93}.

In the limit of a large in-plane magnetic field, the cooperon contribution is strongly suppressed since time reversal symmetry is broken. This can be seen explicitly by computing the lowest transverse eigenvalue $\epsilon_\perp^\pm$ perturbatively in $B$, for both the cooperon and the diffuson contributions. This perturbative approach is valid as long as $R\ll\ell_B$, where $\ell_B$ has to be evaluated for the appropriate in-plane magnetic fields entering the cooperon ($+$) and diffuson ($-$) contributions. (Here, $\ell_B=(1/eB_\parallel)^{1/2}$ denotes the magnetic length.)
Due to the boundary condition of zero normal current, the ground state wavefunction $|{\rm gs}\rangle$ of Eq.\ (\ref{DiffusonEquation}) at zero $B_\parallel$ is a constant with zero transverse eigenvalue. Thus, the leading correction to the eigenvalue is given by
\begin{eqnarray}
   \epsilon_\perp &=& \left\langle {\rm gs}\left| \frac{De^2(B_\parallel)^2}{4}(x^2+y^2)\right|{\rm gs}\right
   \rangle \nonumber\\
    &=& \frac{D}{8\ell_B^2}\left(\frac{R}{\ell_B}\right)^2,
\end{eqnarray}
and we find that
\begin{equation}
  \frac{\epsilon_\perp}{E_c} = \frac{1}{32\pi^2}\left(\frac{LR}{\ell_B^2}\right)^2.
  \label{fracEE}
\end{equation}
For the cooperon contribution, the magnetic field is of the order of twice the applied magnetic field. Thus, by Eq.\ (\ref{ZeroTemperature}) this contribution is exponentially suppressed once the relevant in-plane field is larger than one flux quantum penetrating the ring.

We first focus on the typical persistent current at zero temperature. In this case, the effective in-plane field vanishes for the diffuson contribution, while it strongly suppresses the cooperon contribution. Thus, assuming from now on that $B_\parallel$ is sufficiently large to make $\epsilon_\perp(2B)\gg E_c$, we need to retain only the diffuson contribution and obtain
\begin{eqnarray}
\langle I(\phi)I(\phi')\rangle = \frac{3E_c^2}{\pi^4\phi_0^2}\sum_{p=1}^\infty \frac{1}{p^3} \cos(2\pi p[\varphi-\varphi']).
\label{IIhighField}
\end{eqnarray}
Comparing with Eqs.\ (\ref{HarmonicContent}) and (\ref{RandomPhase}), we find
\begin{equation}
  \langle[I_p^{(+)}]^2\rangle = \langle[I_p^{(-)}]^2\rangle = \frac{1}{2}\langle I_p^2\rangle  = \frac{3E_c^2}{\pi^4\phi_0^2}\frac{1}{p^3}
\end{equation}
for the harmonics of the persistent current.

\begin{figure}
        \includegraphics[width=0.85\columnwidth]{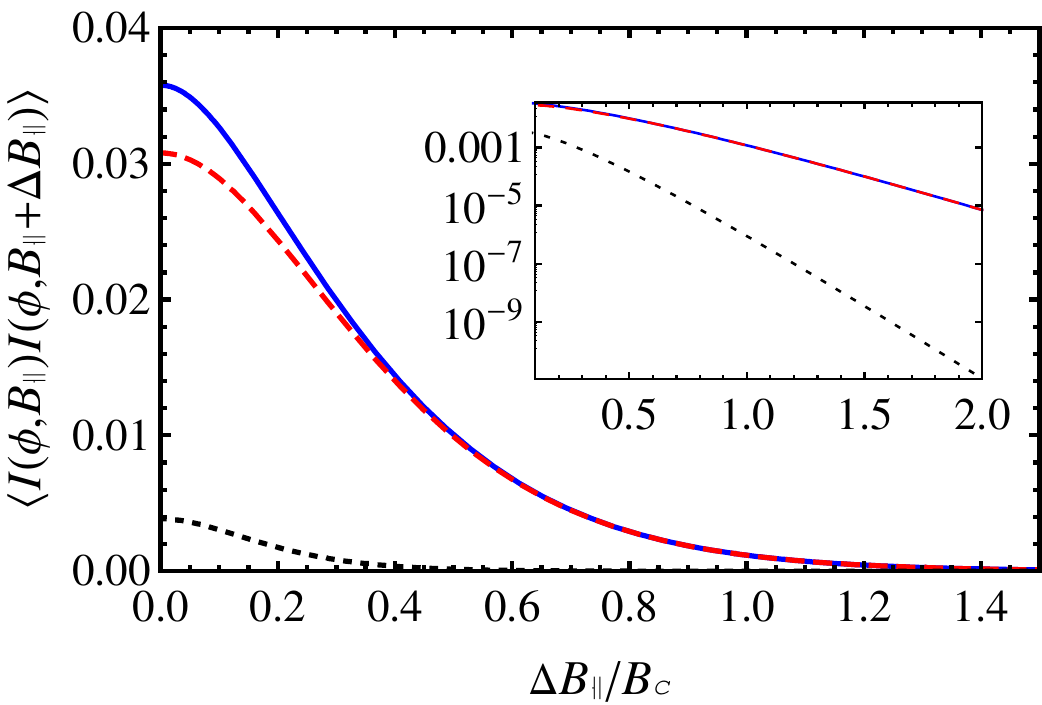}
\caption{(Color online) Current-current correlation function $\langle I(\phi,B_\parallel)I(\phi,B_\parallel+\Delta B_\parallel)\rangle$ (in units of $(E_c/\phi_0)^2$) at zero temperature (solid line). The dashed and dotted lines correspond to the contributions from the first and the second harmonics, respectively. The inset shows the same curves but plotted logarithmically along the vertical axis.
\label{CorrelationFunction}}
\end{figure}

Equation (\ref{ZeroTemperature}), (\ref{functioFp}), and (\ref{fracEE}) also imply that the correlation function of the persistent current at different values of the in-plane magnetic fields falls off exponentially with the magnetic-field difference once the in-plane field changes by more than a flux quantum through the cross section of the ring, i.e., on the scale of the correlation field
\begin{equation}
  B_c = \frac{\sqrt{2}}{\pi}\frac{\phi_0}{LR}.
  \label{CorrelationField}
\end{equation}
Note that the functional dependence of the correlation field on $L$ and $R$ remains the same for much more general field configurations but that the numerical prefactor in Eq.\ (\ref{CorrelationField}) is specific to the toroidal-field model. A plot of the correlation function $\langle I(\phi,B_\parallel) I(\phi,B_\parallel + \Delta B_\parallel)\rangle$ is shown in Fig.\ \ref{CorrelationFunction}. Its exponential fall-off has important ramifications in experiment.  The decay of the correlation function implies that measurements of the persistent current at in-plane fields which are significantly separated from each other on the scale set by $B_c$ are statistically independent. We are thus led to the ergodic hypothesis that averaging over a sufficiently wide range of in-plane fields is equivalent to averaging over the disorder ensemble. This observation is particularly pertinent in view of the novel technique of measuring persistent currents employed in Ref.\ \onlinecite{Harris09} which allows one to obtain the persistent current over a wide range of in-plane magnetic fields.

\begin{figure}[!t]
                \includegraphics[width=0.85\columnwidth]{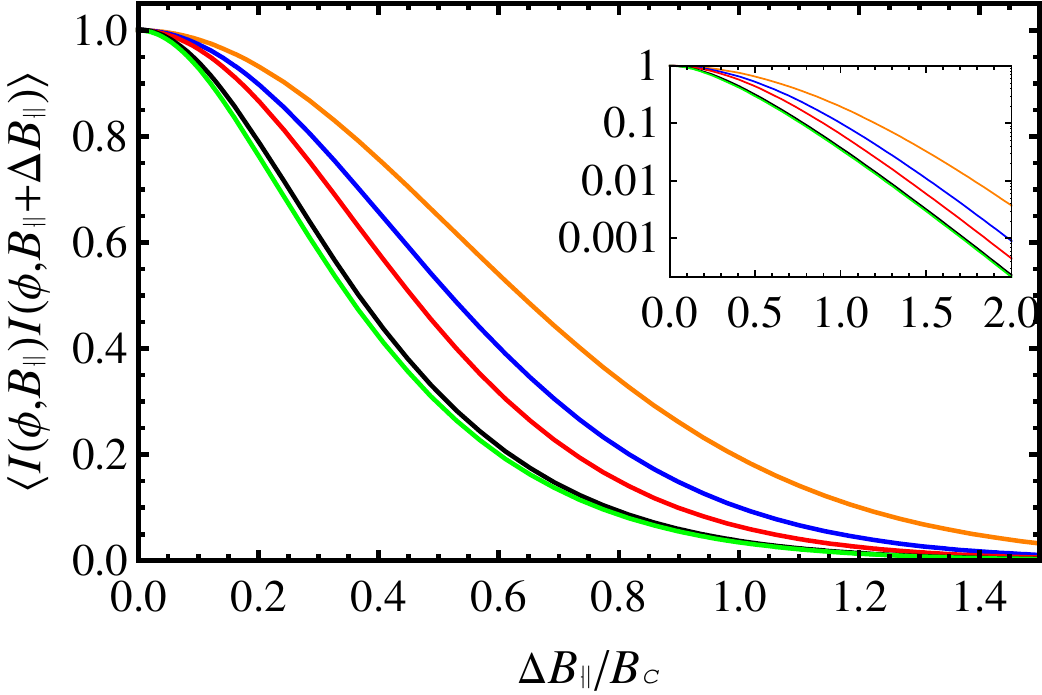}
\caption{(Color online) Current-current correlation function $\langle I(\phi,B_\parallel)I(\phi,B_\parallel+\Delta B_\parallel)\rangle$ (in units of $(E_c/\phi_0)^2$) vs.\ $\Delta B_\parallel$ at
finite temperatures. The curves are normalized to their value at $\Delta B_\parallel=0$ and correspond to $T = 0.01, 0.02, 0.1, 0.2, 0.5 \times E_c$ (from bottom to top). The inset shows the same curves but plotted logarithmically along the vertical axis.
\label{FiniteTemperaturePC}}
\end{figure}

\begin{figure}[b]
                \includegraphics[width=0.85\columnwidth]{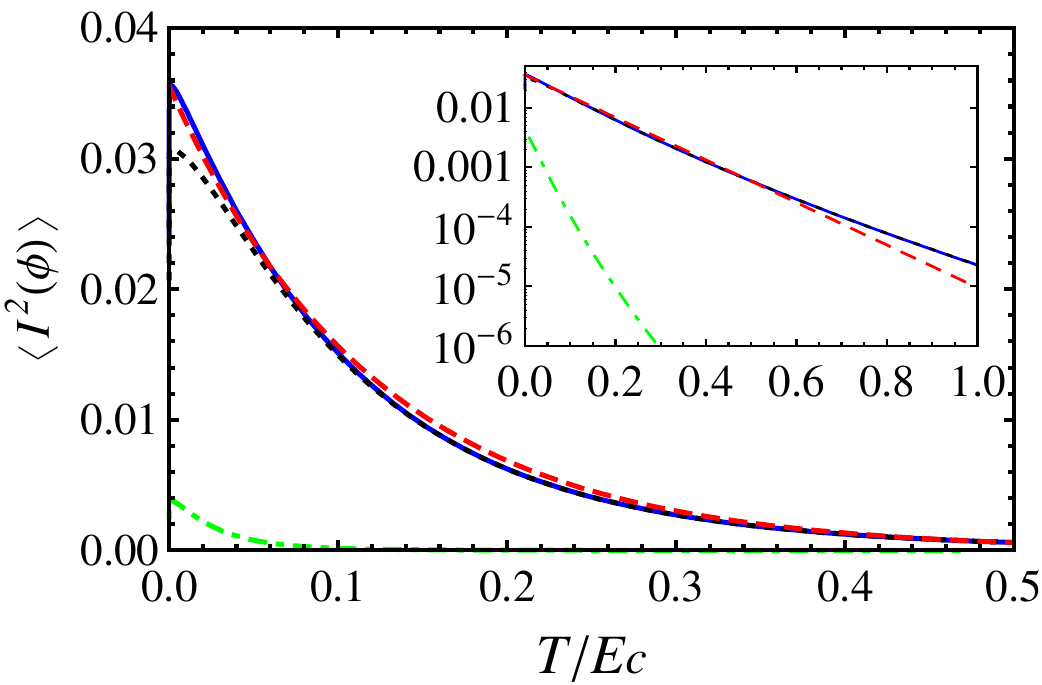}
\caption{(Color online) Temperature dependence of the typical current $\langle I^2(\phi)\rangle$ (blue line). The dependence can be well fitted by an exponential $\langle I^2\rangle \approx c(E_c/\phi_0)^2\exp(-\alpha T/E_c)$ with $c=0.036$ and $\alpha = 8.2$ as shown by the red (dashed) line. The dotted and dash-dotted lines correspond to the contributions from the first and the second harmonics, respectively. The inset shows the same curves but plotted logarithmically along the vertical axis.
\label{TemperatureDependence}}
\end{figure}

We close this section by discussing the temperature dependence of the persistent current at large in-plane magnetic fields. At finite temperatures, the persistent current correlation function depends on $\Delta B$ and temperature $T$ via the two dimensionless variables, $\Delta B/B_c$ and $T/E_c$. The correlation function can be readily evaluated
by combining Eq.\ (\ref{FiniteTemperature}) with Eq.\ (\ref{ZeroTemperature}). Performing the remaining integral numerically, we obtain the results shown in Fig.\ \ref{FiniteTemperaturePC} for the current-current correlation function and in Fig.\ \ref{TemperatureDependence} for the temperature dependence of the typical current. We see from Fig.\ \ref{TemperatureDependence} that the temperature dependence can be approximated as exponential with reasonable (though uncontrolled) accuracy. (Numerical values of the fit are quoted in the figure caption.) Moreover, we observe that the typical persistent current becomes rapidly dominated by the first harmonic as temperature increases.

\subsection{Effects of spin}
\label{spin}

In weak magnetic field and in the absence of spin-orbit scattering, spin enters the persistent current simply through a degeneracy factor of two. Thus, Eq.\ (\ref{IInoSpin}) is modified into
\begin{equation}
\langle I(\phi)I(\phi')\rangle = \frac{24E_c^2}{\pi^4 \phi_0^2}\sum_{p=1}^\infty \frac{1}{p^3}\sin(2\pi p \varphi)\sin(2\pi p \varphi').
\end{equation}
This result includes both the diffuson and cooperon contributions.

In a large applied magnetic field, but still without spin-orbit scattering, the cooperon contribution is suppressed and we have to take the Zeeman energy into account. The corresponding spinless result was given in Eq.\ (\ref{IIhighField}). We can include the spin and Zeeman energies by writing the persistent current as a sum of the contributions of spin-up and spin-down electrons, $I=I_\uparrow+I_\downarrow$. Once the Zeeman energy becomes large compared to the Thouless energy, there are no correlations between $I_\uparrow$ and $I_\downarrow$ and as a result, we find
\begin{equation}
\langle I(\phi)I(\phi')\rangle = \frac{6E_c^2}{\pi^4 \phi_0^2}\sum_{p=1}^\infty \frac{1}{p^3}\cos(2\pi p [\varphi- \varphi']).
\label{ZeroTemperatureII}
\end{equation}

The recent precision measurements\cite{Harris09} of the persistent current were performed on samples whose spin-orbit scattering length is smaller than or of order of the circumference of the rings, as deduced from weak-localization measurements. For this reason, we now turn to a more thorough discussion of the consequences of the electron spin, which in addition accounts for the spin-orbit scattering. This can be done by a standard extension of the diagrammatic technique for diffusive systems.\cite{Vavilov03} To be specific, we focus on sufficiently large magnetic fields that the cooperon no longer contributes significantly. Extensions to include the cooperon contribution at weak fields would pose no additional complications.

\begin{figure}[b]
        \includegraphics[width=0.85\columnwidth]{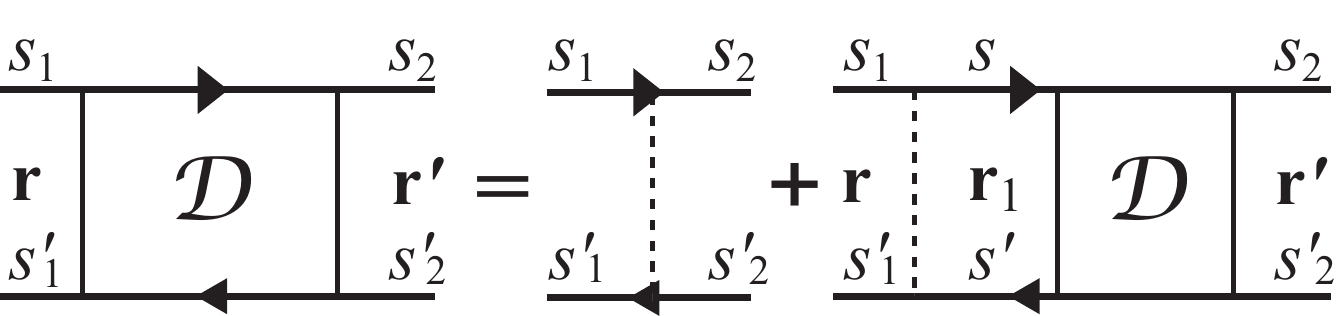}
\caption{Diagrammatic representation of the equation of motion for the diffuson ${\cal D}_{s^\prime_1 s^\prime_2}^{s_1 s_2}({\bf r},{\bf r}', \epsilon)$. Full lines represent electronic Green functions and dashed lines denote disorder and spin-orbit scattering.
\label{DiffusonDiagram}}
\end{figure}

Including spin indices, we define the diffuson ${\cal D}_{s^\prime_1 s^\prime_2}^{s_1 s_2}({\bf r},{\bf r}', \epsilon)$ as shown in Fig.\ \ref{DiffusonDiagram} and view it as a $4\times4$-matrix ${\bf D}({\bf r},{\bf r}', \epsilon)$ where $(s_1, s_1^\prime)$ labels the rows and $(s_2, s_2^\prime)$ the columns. With the ordering $(s,s')=(\uparrow\uparrow, \uparrow\downarrow, \downarrow\uparrow, \downarrow\downarrow)$, one obtains the equation\cite{Vavilov03}
\begin{widetext}
\begin{equation}
   \left[-D\left(\nabla-{ie}{\bf A_-}\right)^2 +i\epsilon +H_Z+H_{\rm so}\right]{\bf D}({\bf r},{\bf r}',
   \epsilon) = \frac{1}{2\pi N(0)\tau^2} \delta({\bf r}-{\bf r}')
\end{equation}
by the standard procedure, starting with the diagrammatic representation shown in Fig.\ \ref{DiffusonDiagram}. ($N(0)$ denotes the density of states at the Fermi energy and $\tau$ is the elastic scattering time.) Here, the contribution of the Zeeman energy $E_Z$ yields the term
\begin{equation}
 H_Z = \left[\begin{array}{cccc} 0 &0 & 0 & 0 \\
 0 & -2iE_Z & 0 & 0 \\
 0 & 0 & 2iE_Z & 0 \\
 0 & 0 & 0 & 0 \end{array}\right],
\end{equation}
while spin-orbit scattering is included through
\begin{equation}
 H_{\rm so} = \frac{2}{3\tau_{\rm so}}\left[\begin{array}{cccc} 1 &0 & 0 & -1 \\
 0 & 2 & 0 & 0 \\
 0 & 0 & 2 & 0 \\
 -1 & 0 & 0 & 1 \end{array}\right]
\end{equation}
in terms of the spin-orbit scattering time $\tau_{\rm so}$.

By retracing the steps leading up to Eq.\ (\ref{CurrentThroughDifusion}) in the presence of spin effects, we obtain for the correlation function of the persistent current,
\begin{eqnarray}
 C_I^{(0)}(E,B;E',B') = -\frac{1}{4\pi^2}\frac{\partial^2}{\partial\phi\partial \phi'} \int_0^\infty
  d\sigma {\rm Im} \int_{-2\sigma}^{2\sigma} d\epsilon\, \epsilon
  {\rm Tr}\left(\frac{1}{-D[\nabla-{ie}{\bf A}_-]^2 + i(\epsilon+E-E') + H_Z + H_{\rm so}}\right),
\label{CurrentThroughDifusionWithSpin}
\end{eqnarray}
where ${\rm Tr}$ now denotes a trace over configuration space and the four-dimensional spin space.

In the limit of large Zeeman splitting, $E_Z \gg E_c$, the modes $\uparrow\downarrow$ and $\downarrow\uparrow$ are exponentially suppressed. For negligible spin-orbit scattering, we then obtain two massless modes $\uparrow\uparrow\pm \downarrow\downarrow$. As a result, the correlation function is twice larger than the result for spinless electrons given in Eq.\ (\ref{ZeroTemperature}), in agreement with Eq.\  (\ref{ZeroTemperatureII}). As the spin-orbit scattering increases, only the density mode $\uparrow\uparrow + \downarrow\downarrow$ remains massless and in the limit of strong spin-orbit scattering, we recover the result in Eq.\ (\ref{ZeroTemperature}) for spinless electrons.

More generally, we can discuss the crossover between the limits of weak and strong spin-orbit scattering rate. One finds
\begin{eqnarray}
 &&C_I^{(0)}(E,B;E',B') = -\frac{1}{4\pi^2}\frac{\partial^2}{\partial\phi\partial \phi'} \int_0^\infty
  d\sigma {\rm Im} \int_{-2\sigma}^{2\sigma} d\epsilon\, \epsilon
  \nonumber\\
  &&  \,\,\,\,\,\,\,\,
  \times{\rm Tr}\left(\frac{1}{-D[\nabla-{ie}{\bf A}_-]^2 + i(\epsilon+E-E')}
  +  \frac{1}{-D[\nabla-{ie}{\bf A}_-]^2 + i(\epsilon+E-E') + \frac{4}{3\tau_{\rm so}}}\right.
  \nonumber\\
  && \,\,\,\,\,\,\,\,
  + \left.\frac{1}{-D[\nabla-{ie}{\bf A}_-]^2 + i(\epsilon+E-E'+2E_Z) + \frac{4}{3\tau_{\rm so}}}
  +\frac{1}{-D[\nabla-{ie}{\bf A}_-]^2 + i(\epsilon+E-E'-2E_Z) + \frac{4}{3\tau_{\rm so}}}\right) ,
\end{eqnarray}
where the trace is now over configuration space only. Specifying again to the toroidal-field model, we obtain
\begin{eqnarray}
  &&C_I^{(0)}(E,B;E',B') = -\frac{8E_c^2}{\pi}\frac{\partial^2}{\partial\phi\partial \phi'}\sum_{p=1}^\infty
  \cos(2\pi p\varphi_\pm)  \left[F_p\left(i\frac{E-E'}{E_c}+ \frac{\epsilon^-_\perp}{E_c}\right) \right.
  \nonumber\\
  &&\,
  \left.  +
  F_p\left(i\frac{E-E'}{E_c}+\frac{\epsilon^-_\perp+ \frac{4}{3\tau_{\rm so}}}{E_c}\right)
  +
  F_p\left(i\frac{E-E'+2E_Z}{E_c}+\frac{\epsilon^-_\perp+ \frac{4}{3\tau_{\rm so}}}{E_c}\right)
  +
  F_p\left(i\frac{E-E'-2E_Z}{E_c}+\frac{\epsilon^-_\perp+ \frac{4}{3\tau_{\rm so}}}{E_c}\right)
  \right]
  \label{cross_over}
\end{eqnarray}
\end{widetext}
where the function $F_p(z)$ had been defined in Eq.\ (\ref{functioFp}).

\begin{figure}[b]
        \includegraphics[width=0.85\columnwidth]{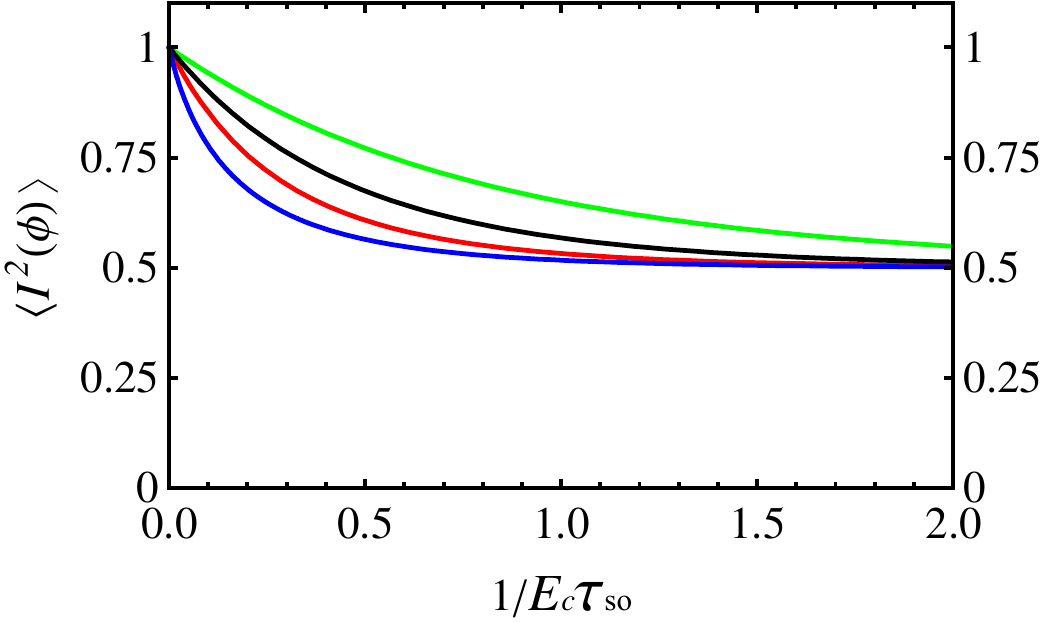}
\caption{(Color online) Crossover of the typical current $\langle I^2(\phi)\rangle$ as function of the spin-orbit scattering rate. The curves, corresponding to temperatures $T = 0.01, 0.1, 0.3, 1.0 \times E_c$ (from bottom to top),  are normalized to the value of $\langle I^2(\phi)\rangle$ in the limit of vanishing spin-orbit scattering rate. All curves are plotted in the limit of large in-plane field where the cooperon contribution is suppressed and the Zeeman energy is large compared to the Thouless energy.
\label{crossover}}
\end{figure}

Combining Eq.\ (\ref{cross_over}) with Eq.\ (\ref{FiniteTemperature}) and setting $\epsilon^-_\perp=0$, we can obtain the crossover of the typical current between the limits of weak and strong spin-orbit scattering for arbitrary temperature. (Note that the results for the typical current are not restricted to the toroidal-field model.) Corresponding numerical results in the limit of large Zeeman splitting (where the last two terms in the square bracket in Eq.\ (\ref{cross_over}) can be neglected) are plotted in Fig.\ \ref{crossover}, which show that the crossover becomes slower as temperature increases.

\section{Interaction contribution}
\label{interaction}

We now turn to a discussion of the interaction-contribution to the persistent current in high magnetic fields.
Adapting the first-order correction in the interaction $V$ derived in Ref.\ \onlinecite{Ambegaokar90} to the case of a finite magnetic field, one finds for the disorder-averaged contribution to the grand canonical potential
\begin{eqnarray}
\label{IntGrandPot-1}
\Delta \Omega &=&  \frac{N(0)\bar{V}}{\pi}\int_0^{\infty} dE \coth
(\frac{E}{2T})E \nonumber \\
&&\times {\rm Re}{\rm Tr}
\frac{1}{-D(\nabla-{2ie}{\bf A})^2+iE}
\end{eqnarray}
Here, $\bar{V}$ is the Fourier component of the screened Coulomb interaction potential averaged in momentum space.\cite{Ambegaokar90} In a field much stronger than the upper critical field of the ring we may constrain considerations to the lowest-order correction, Eq.~(\ref{IntGrandPot-1}). To estimate the interaction contribution to the average persistent current, we again employ the toroidal-field model introduced in Sec.\ \ref{independent}. Then, the eigenvalue problem and boundary conditions for the cooperon here are identical to those in Eqs.\ (\ref{BoundaryCondition}) and (\ref{diffuson}), respectively.

We denote the cooperon eigenvalues by $\epsilon^{(l)}_{n,m,\phi}$ with $l,n,m$ being the radial, longitudinal and azimuthal quantum numbers, respectively.  Due to cylindrical symmetry the cooperon modes can be found by separation of variables, with the replacement $n \rightarrow n-2\phi/\phi_0$ added to take into account the Aharonov-Bohm flux. In
distinction from Sec.\ \ref{independent}, the vector potential ${\bf A}$ in Eq.\ (\ref{IntGrandPot-1}) corresponds to the total field so that $\ell_B \ll R$. In this limit, the radial equation can be approximated to lowest order in $\ell_B/R$ as
\begin{equation}\label{IntEigenEq}
\frac{D}{\ell_B^2}\left( -\frac{\partial^2}{\partial x^2} + (\kappa_m-x)^2+\ell_B^2 k_{n,\phi}^2\right)\chi(x)=\epsilon^{(l)}_{n,m,\phi}\chi(x)
\end{equation}
where $x=r/\ell_B$ is a scaled distance from the center of the cross section, $\kappa_m=m \ell_B/R$ and
$k_{n,\phi}=2\pi(n-2\phi/\phi_0)/L$.  Note that the ratio between the radial and the longitudinal terms in Eq.\  (\ref{IntEigenEq}) is dominated by $L/\ell_B$. The eigenvalues can be written as
\begin{equation}
\epsilon^{(l)}_{n,m,\phi}=D\left(\frac{2\pi}{L}\right)^2\left[ (n-\frac{\phi}{\phi_0/2})^2+\left(\frac{L}{2\pi \ell_B}\right)^2\lambda_l(\kappa_m)
\right]
\end{equation}
where the values of $\lambda_0(\kappa_m)$ for the lowest branch of eigenstates ($l=0$) can be estimated by using the variational method with a Gaussian trial solution. The function $\lambda_0(\kappa)$ has a shallow minimum $\lambda_0^*=(1-2/\pi)^{1/2}$ at $\kappa_m^*=(\pi^2-2\pi)^{-1/4}$.  Using the eigenvalues $\epsilon^{(l)}_{n,m,\phi}$ to evaluate the trace in Eq.~(\ref{IntGrandPot-1}), it is straightforward to show that the contribution to persistent current $\Delta I=-\partial \Delta \Omega /\partial \phi$ is periodic in $\phi\rightarrow \phi+\phi_0/2$, and for $T=0$ can be written as
\begin{equation}
\Delta I=\frac{N(0)\bar{V}}{\pi} \left( \frac{2\pi}{L}\right)^2\frac{4hD}{\phi_0}\sum_{p=1}^{\infty} p g_p \sin\left( 2\pi p \frac{\phi}{\phi_0/2}\right)\,.
\end{equation}
In the regime of experimental interest, $L\gg R \gg \ell_B$, the coefficients $g_p$,
\begin{equation}
g_p=\frac{1}{2p^3 \pi^2}\sum_{m=0}^{\infty} e^{-pL/\ell_B \sqrt{\lambda_0(\kappa_m)}}\left(1+ p \frac{L}{\ell_B}\sqrt{\lambda_0(\kappa_m)}\right)
\end{equation}
can be estimated by evaluating the sum in the saddle-point approximation,
\begin{equation}
g_p\approx 0.13p^{-3.5}\left( \frac{R}{\sqrt{\ell_B L}}\right) \left[1+p\frac{L}{\ell_B}\sqrt{\lambda_0^*}\right]e^{-\frac{L}{\ell_B}p\sqrt{\lambda_0^*}}\,.
\end{equation}
All harmonics of the average persistent current are exponentially suppressed; the higher the harmonic $p$, the stronger is the suppression. Note that this implies that for sufficiently strong magnetic field, measurements of the average current, e.g., by employing large arrays of rings, should be dominated by the canonical-ensemble contribution of the free-electron model.\cite{Altshuler91, Schmid91, Oppen91}

\begin{figure}[t]
        \includegraphics[width=0.85\columnwidth]{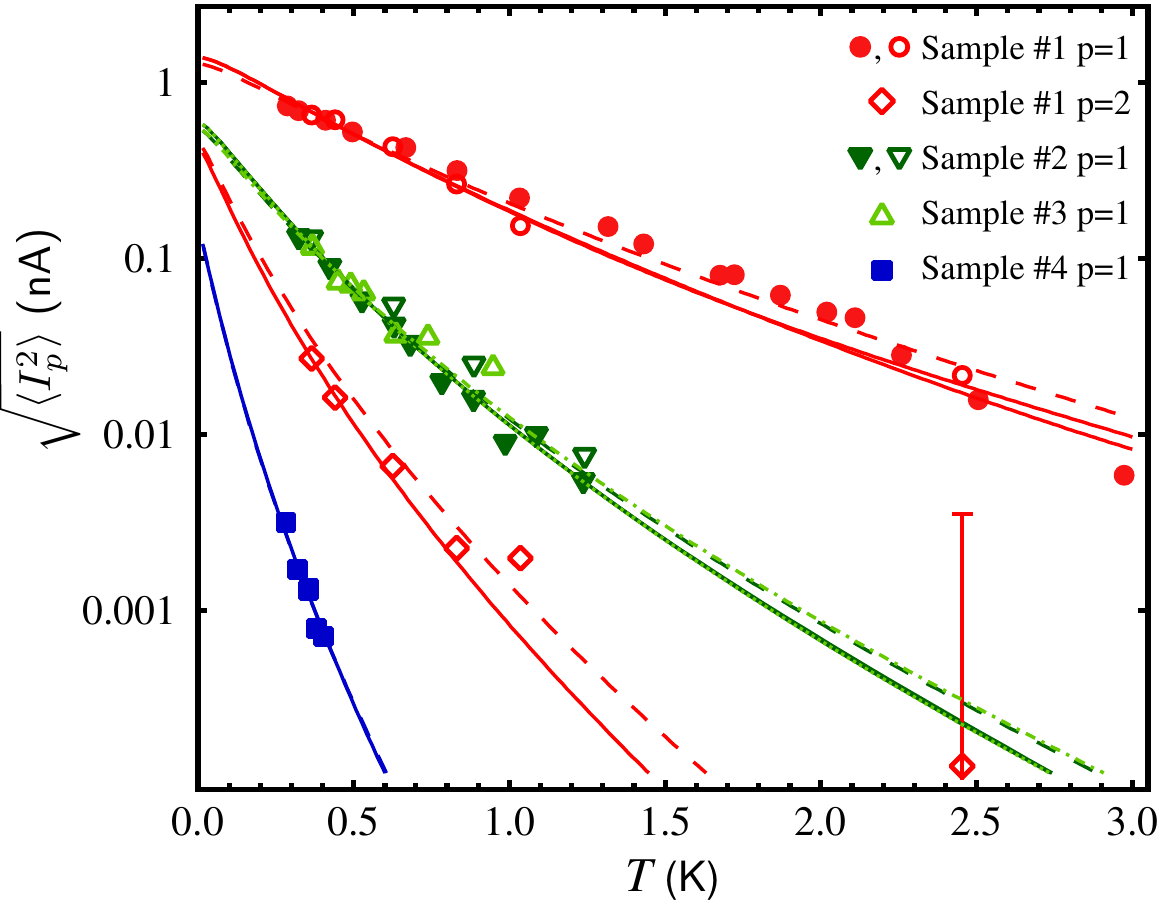}
\caption{(Color online) Temperature dependence of the typical current contribution from the $p^{\rm th}$ harmonic $\sqrt{\langle I_p^2(T) \rangle}$.  The markers are the data first presented in Ref.\ \onlinecite{Harris09}. The solid curves represent new fits to the data using Eqs.\ (\ref{FiniteTemperature}) and (\ref{cross_over}) while the dashed curves show the fits from Ref.\ \onlinecite{Harris09}. The sample parameters and best-fit parameters are given in Table \ref{diffusionconstants}.  Closed and open markers denote measurements taken during different cooldowns, over different field ranges, and at different magnetic field orientations.  In the case of the $p=1$ data from Sample \#1, the two different field ranges over which the closed and open markers were taken lead to slightly different values of the fitting function at high temperature, with the lower curve corresponding to the closed markers and the upper curve to the open markers.  The new fit curves for Samples \#2 \& \#3 are indistinguishable, as are the old and new fit curves for Sample \#4.
\label{datafit}}
\end{figure}

\section{Comparison to Experiment}
\label{datacomparison}

The recent development of cantilever-based torsional magnetometers with integrated mesoscopic rings\cite{Bleszynski08} resulted in measurements of the rings' persistent current in the presence of large magnetic fields.\cite{Harris09} Here we briefly review these measurements and compare them with the calculations from the preceding sections. This comparison is most readily performed by fitting the measured temperature dependence of the current to the form predicted in Eqs.\ (\ref{FiniteTemperature}) and (\ref{cross_over}).

\begin{table*}
\caption{Sample parameters. ``Marker'' refers to the markers used in Fig.\ \ref{datafit}, with closed and open markers representing two different cooldowns of the same sample.  For the closed markers the angle between the magnetic field and the plane of the rings was 6$^{\circ}$ and $T_0=323$ mK, while for the open markers the angle was 45$^{\circ}$ and $T_0=365$ mK.  $N$ denotes the number of rings in the sample.  The ring circumference and linewidth are given by $L$ and $w$.  The thickness of each sample was 90 nm.  The spin orbit scattering length $L_{\rm so}=1.1\: \pm\: 0.25 \:\mu \mathrm{m}$.  $B_{\rm min}$ and $B_{\rm max}$ give the bounds for measurements of $I(B)$ taken over smaller field ranges.  $D_L$ and $D_{\rm ZSO}$ are extracted from fitting the
persistent current data. $D_L$ is the best-fit value of the diffusion constant found in Ref.\ \onlinecite{Harris09}, which assumed the limit of strong spin-orbit scattering and large Zeeman splitting.  $D_{\rm ZSO}$ is the best-fit value of the diffusion constant found by taking into account the finite spin-orbit scattering rate and Zeeman splitting as described in Section\ \ref{datacomparison}.  The estimated uncertainty in all fit coefficients is 6\%. }
                \begin{tabular*}{1.7\columnwidth}{@{\extracolsep{\fill}} | c c c c c c c c c c |}\hline
                        Sample  & Marker &      $p$     &       $N$     & $L$ ($\mu$m) & $w$ (nm) & $B_{\rm min}$ (T) & $B_{\rm max}$ (T) & $D_{L}$ (cm$^2$/s) & $D_{\rm ZSO}$ (cm$^2$/s) \\     \hline

                        \multirow{3}{*}{\#1} & \includegraphics{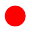} & 1 & \multirow{3}{*}{1680} & \multirow{3}{*}{1.9} & \multirow{3}{*}{115} & 6.2 & 6.8 & \multirow{3}{*}{271} & \multirow{3}{*}{234} \\
                         & \includegraphics{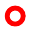} & 1 & & & & 5.0 & 5.2 & & \\
                         & \includegraphics{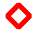} & 2 & & & & 5.0 & 5.2 & & \\     \hline

                        \multirow{2}{*}{\#2} & \includegraphics{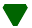} & \multirow{2}{*}{1} & \multirow{2}{*}{990} & \multirow{2}{*}{2.6} & \multirow{2}{*}{85} & 7.15 & 7.60 & \multirow{2}{*}{214} & \multirow{2}{*}{195} \\
                         & \includegraphics{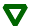} & & & & & 5.39 & 5.48 & & \\    \hline

                        \#3 & \includegraphics{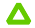} & 1 & 1 & 2.6 & 85 & 8.32 & 8.40 & 215 & 195 \\  \hline

                        \#4 & \includegraphics{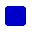} & 1 & 242 & 5.0 & 85 & 7.1 & 7.3 & 205 & 196 \\  \hline
                \end{tabular*}

        \label{diffusionconstants}
\end{table*}

The parameters characterizing each sample are collected in Table\ \ref{diffusionconstants}. The temperature dependence of the $p^{\rm th}$ harmonic $\sqrt{\langle I_p^2\rangle}$ of each  sample's typical current was determined as follows. At a single temperature $T_0$, the mean square amplitude of the $p^{\rm th}$ harmonic of the current was extracted from a measurement of $I(B)$ taken over a range of $B$ spanning many $B_c$. This large span ensured that the mean was determined from a large number of independent measurements, as discussed at the end of Sec.\ \ref{ToroidalMagneticField}. For each sample, the form of $I(B)$ was found to be independent of temperature except for an overall scaling. This scaling was determined by measuring $I(B)$ over a smaller field range (with bounds denoted by $B_{\rm min}$ and $B_{\rm max}$) at each subsequent temperature and comparing the magnitude of each harmonic with the value measured over the same field range at $T_0$. This procedure, as well as other details of the measurements, are described in detail in Ref.\ \onlinecite{Harris09}.  The resulting values of $\sqrt{\langle I_p^2\rangle}$ are shown in Fig.\ \ref{datafit}.

In Ref.\ \onlinecite{Harris09}, this data was analyzed by assuming the limit of strong spin orbit scattering: $1/\tau_{\rm so}\gg \{E_c, T\}$, and large Zeeman splitting, $E_Z\gg \{E_c,T\}$. As can be seen from the sample parameters listed in Table\ \ref{diffusionconstants}, this assumption is fairly accurate though not exact.  For these samples $0.075 < 1/E_c \tau_{\rm so} < 0.47$ while $0.15 <  T/E_c < 1.7$.  From Fig.\ \ref{crossover} it is clear that these parameters are not fully within the strong spin orbit scattering limit. Additionally, for the smallest rings (Sample \#1) the limit of large Zeeman splitting does not hold at the highest temperatures in Fig.\ \ref{datafit} where $ T/E_Z \approx 0.36$ and deviations from the large Zeeman splitting limit change $\sqrt{\langle I_p^2\rangle}$ by as much as 5\%. For Samples \#2, \#3, and \#4 $ T/E_Z < 0.17$ resulting in  $< 1\%$ deviations of $\sqrt{\langle I_p^2\rangle}$  from the large Zeeman splitting limit. As a result the measurements of Ref.\ \onlinecite{Harris09} were not fully in the strong spin orbit scattering, large Zeeman splitting limit, so we reanalyze the data here, taking into account the full dependence of $\sqrt{\langle I_p^2\rangle}$ on spin orbit scattering and Zeeman splitting.

We fit the data from Ref.\ \onlinecite{Harris09} (Fig.\ \ref{datafit}) using the expression for $\sqrt{\langle I_p^2 (T,D,L_{\rm so},E_z)\rangle}$ derived from Eqs.\ (\ref{FiniteTemperature}) and (\ref{cross_over}). The only fitting parameter is the electron diffusion constant $D$.  The spin orbit length $L_{\rm so}\equiv\sqrt{D \tau_{\rm so}} = 1.1\: \pm\: 0.25\: \mu \mathrm{m}$ was determined independently from magnetotransport measurements of a wire codeposited with the rings.\cite{Harris09}  Since each data point in Fig.\ \ref{datafit} is extracted from measurements of $I(B)$ made over a range of $B$, we cannot use a single value of the Zeeman splitting; instead, we average over the magnetic field range to obtain the fitting function
\begin{eqnarray}
\lefteqn{\sqrt{\langle I_p^2 (T,D,L_{\rm so},B_{\rm min},B_{\rm max})\rangle} = } \nonumber \\
& & \sqrt{\frac{\int_{B_{\rm min}}^{B_{\rm max}}dB\langle I_p^2 (T,D,L_{\rm so},E_Z(B))\rangle}{B_{\rm max}-B_{\rm min}}}.
\label{EZaverage}
\end{eqnarray}
The best-fit values of the diffusion constant $D_{\rm ZSO}$ are given in Table\ \ref{diffusionconstants}.  The corresponding fits are shown in Fig.\ \ref{datafit} as solid lines.  For comparison the values of the diffusion constant found in Ref.\ \onlinecite{Harris09}, $D_L$, are also given in Table\ \ref{diffusionconstants} and the corresponding fits are shown as dashed lines in Fig.\ \ref{datafit}.

Figure\ \ref{datafit} and Table\ \ref{diffusionconstants} show that the finiteness of the spin-orbit scattering rate and the Zeeman energy result in small but noticeable changes to the fitted curves and the extracted values of $D$. We find that most of the difference is due to the finite spin orbit scattering rate, which leads to a non-negligible contribution to the current from the second $F_p$ term in Eq.\ (\ref{cross_over}).

The finite Zeeman energy modifies the current via the last two $F_p$ terms in Eq.\ (\ref{cross_over}), leading to a correction which becomes appreciable ($> 1 \%$) only for the higher temperature measurements of Sample \#1. The resulting correction oscillates as a function of temperature, resulting in a best-fit value of $D$ indistinguishable from the case of large Zeeman splitting.

The values of $D_{\rm ZSO}$ for Samples \#2, \#3, and \#4 agree with each other to within the experimental uncertainty (which is estimated to be $6 \%$ in the Supplemental Online Material of Ref.\ \onlinecite{Harris09}). This agreement is consistent with the fact that the rings in these three samples have the same cross-sectional dimensions.  The value of $D_{\rm ZSO} = 234 \:\mathrm{ cm}^2/\mathrm{s}$ measured for Sample \#1 is somewhat larger, which may reflect these rings' larger cross section. Resistivity measurements of the codeposited wire having the same cross section as Sample \#1 give $D = 260 \pm 12 \:\mathrm{ cm}^2/\mathrm{s}$, consistent with the value measured for Sample \#1.\cite{Harris09}

\section{Conclusions}
\label{conclusions}

Motivated by a new and highly sensitive experimental technique \cite{Harris09} for measuring mesoscopic persistent currents, we presented a theory of persistent currents in large, but non-quantizing, magnetic fields. The theoretical results of this paper formed the basis for establishing the remarkable quantitative agreement between experiment and theory found in Ref.\ \onlinecite{Harris09} and further refined in Sec.\ref{datacomparison}. To reach this agreement, we not only needed to take into account the large magnetic field, both for the single-particle and the interaction contributions to the persistent current, but also spin effects.

In addition to forming the basis for a quantitative comparison with experiment, it is also worth emphasizing several theoretical conclusions from our results.

(i) The magnetic field penetrating the ring leads to qualitative changes in the dependence of the persistent current on the Aharonov-Bohm flux. At zero magnetic field, the persistent current is a periodic function of flux. Zero flux as well as integer and half-integer multiples of the flux quantum are special points where the persistent current vanishes. At large magnetic fields, the persistent current $I(\phi)$ is still a periodic function of flux, but the typical magnitude $\langle I^2(\phi)\rangle $ is no longer dependent on flux.

(ii) Previous theoretical works have shown that there are two principal contributions to mesoscopic persistent currents: a free-electron contribution and an interaction contribution. In experiments, it is not always easy to disentangle these two contributions (especially for the even harmonics of the persistent current). In fact, while the interaction contribution is expected to dominate the ensemble-averaged persistent current, both of them contribute significantly in single- or few-ring experiments. We conclude from our results that the application of a large magnetic field penetrating the ring strongly suppresses the interaction contribution to the persistent current so that the technique of Ref.\ \onlinecite{Harris09} provides direct access to the free-electron contribution.

(iii) One of the principal advantages of the experimental technique of Ref.\ \onlinecite{Harris09} is that unlike SQUID-based approaches, it allows for measurements over a wide range of magnetic fields and thus of many oscillations of the persistent current with flux. Our results for the autocorrelation function of the persistent current at different magnetic fields imply that averaging over magnetic field is equivalent to an ensemble average (ergodic hypothesis). One of the possibilities raised by this result is a direct measurement of the entire distribution function of the persistent current.

The experimental technique of Ref.\ \onlinecite{Harris09} has brought many additional experiments on persistent currents and related phenomena within experimental reach. Our approach should be a valuable starting point for analyzing such future experiments.

\begin{acknowledgments}
This work was supported in part by DOE grant DE-FG02-08ER46482 (LG), by DIP (FvO), as well as by NSF grants 0706380 and 0653377 (JGEH). FvO and LG acknowledge the hospitality of KITP while part of this work has been performed.
\end{acknowledgments}

\appendix
\section{Arbitrary magnetic-field configurations}

Within the toroidal-field model discussed and employed in Secs.\ \ref{ToroidalMagneticField} and \ref{interaction}, we could account for the magnetic field penetrating the ring by perturbation theory. This perturbative calculation was valid as long as $R\ll \ell_B$. The toroidal-field model was special in that at the surface of the ring, the vector potential ${\bf A}_\parallel({\bf r})$ associated with the magnetic field penetrating the ring points parallel to the surface. As a result, the vector potential does not enter into the boundary condition Eq.\ (\ref{BoundaryCondition}) for the equation of the cooperon or the diffuson. Then, computing the perturbative shift of the eigenvectors by the in-plane magnetic field amounts to conventional perturbation theory as familiar from quantum mechanics.

This is no longer the case for more general (and more realistic) models of the in-plane field. Instead, the magnetic field enters not only the diffuson or cooperon equation, but also the boundary condition. In this appendix, we show how
one can in principle reduce the resulting generalized problem of perturbation theory to the conventional case of quantum-mechanical perturbation theory.

The equation for the cooperon or the diffuson is given by
\begin{equation}
   -D \left[\nabla - {ie}{\bf A}_\perp - {ie} {\bf A}_\parallel\right]^2\psi = E\psi ,
\end{equation}
with the appropriate choice of magnetic field. This equation needs to be solved in conjunction with the boundary condition
\begin{equation}
 \left.{\bf \hat n}\cdot[\nabla- {ie}{\bf A}_\perp - {ie} {\bf A}_\parallel]\psi\right|_\Sigma =0
\end{equation}
valid at the surface $\Sigma$ of the ring. We make the gauge choice $\nabla \cdot {\bf A}_\parallel = 0$. The basic observation is that we can eliminate the vector potential ${\bf A}_\parallel$ from the boundary condition by the gauge transformation
\begin{equation}
  \psi({\bf r}) = e^{i f({\bf r})} \psi_1({\bf r}).
\end{equation}
The new function $\psi_1({\bf r})$ satisfies the modified diffusion equation
\begin{equation}
   -D \left[\nabla - {ie}{\bf A}_\perp - {ie} {\bf A}_\parallel+ i \nabla f\right]^2\psi_1 = E\psi_1
   \label{DiffusonWithMagneticField}
\end{equation}
with boundary condition
\begin{equation}
 \left.{\bf \hat n}\cdot[\nabla- {ie}{\bf A}_\perp - {ie} {\bf A}_\parallel+ i \nabla f]\psi_1\right|_\Sigma =0.
\end{equation}
If we choose the gauge transformation such that
\begin{equation}
   {e}\left. {\bf \hat n}\cdot{\bf A}_\parallel\right|_\Sigma = \left.{\bf \hat n}\cdot \nabla f\right|_\Sigma
   \label{f1}
\end{equation}
combined with the gauge choice
\begin{equation}
  \nabla^2 f =0,
  \label{f2}
\end{equation}
we reduce the problem to a form which is amenable to standard techniques of perturbation theory, namely Eq.\ (\ref{DiffusonWithMagneticField}) combined with the boundary condition
\begin{equation}
 \left.{\bf \hat n}\cdot[\nabla- {ie}{\bf A}_\perp]\psi_1\right|_\Sigma =0.
\end{equation}
The principal technical difficulty consists in solving the ``electrostatics'' problem defined by Eqs.\ (\ref{f1}) and (\ref{f2}) to find the function $f({\bf r})$.

\end{document}